\documentclass[twocolumn,showpacs,pre,floatfix,amsmath,amssymb,notitlepage]{revtex4-1}
\usepackage{epsfig}
\usepackage{subfigure}
\usepackage{amsmath}
\usepackage{color}
\usepackage{amssymb}
\usepackage{setspace}
\usepackage{graphicx}
\usepackage{dcolumn}
\usepackage{bm}
\usepackage{times}
\usepackage{enumerate}
\usepackage{footnote}
\input epsf

\graphicspath{{figures/}}

\begin{document}

\author{Per Sebastian Skardal}
\email{persebastian.skardal@trincoll.edu} 
\affiliation{Department of Mathematics, Trinity College, Hartford, CT 06106, USA}

\title{Symmetry and symmetry breaking in coupled oscillator communities}

\begin{abstract}
With the recent development of analytical methods for studying the collective dynamics of coupled oscillator systems, the dynamics of communities of coupled oscillators have received a great deal of attention in the nonlinear dynamics community. However, the majority of these works treat systems with a number of symmetries to simplify the analysis. In this work we study the role of symmetry and symmetry-breaking in the collective dynamics of coupled oscillator communities, allowing for a comparison between the macroscopic dynamics of symmetric and asymmetric systems. We begin by treating the symmetric case, deriving the bifurcation diagram as a function of intra- and inter-community coupling strengths. In particular we describe transitions between incoherence, standing wave, and partially synchronized states and reveal bistability regions. When we turn our attention to the asymmetric case we find that the symmetry-breaking complicates the bifurcation diagram. For instance, a pitchfork bifurcation in the symmetric case is broken, giving rise to a Hopf bifurcation. Moreover, an additional partially synchronized state emerges, as well as a new bistability region.
\end{abstract}


\maketitle

\section{Introduction}\label{sec:01}

The dynamics of large systems of network-coupled oscillators represents an important area of research with a wide range of applications in the physical, biological, engineering, and social sciences~\cite{Winfree2001,Strogatz2003,Pikovsky2003,Arenas2008PR}. Specific examples include rhythmic oscillations in populations of fireflies~\cite{Buck1988QRB}, synchronization of cardiac pacemakers~\cite{Glass1988}, mammalian circadian rhythms~\cite{Strogatz1987JMB,Lu2016Chaos}, synchronization of cell cycles~\cite{Prindle2012Nature}, Josephson junction arrays~\cite{Wiesenfeld1998PRE}, and dynamics of power grids~\cite{Rohden2012PRL,Skardal2015SciAdv}. A particularly important model for studying a wide range of phenomena in coupled oscillator systems is the Kuramoto model~\cite{Kuramoto1984}, which consists of $N$ coupled phase oscillators that, when placed on a network, evolve according to
\begin{align}
\dot{\theta}_n=\omega_n+K\sum_{m=1}^NA_{nm}\sin(\theta_m-\theta_n),\label{eq:01}
\end{align}
where $\theta_n$ represents the phase of oscillator $n$ with $n=1,\dots,N$, $\omega_n$ is the natural frequency of $n$, $K$ is the global coupling strength, and the adjacency matrix $A$ encodes the network topology: entries $A_{nm}$ represent the (possibly weighted) connection between oscillators $n$ and $m$. The relationship between the macroscopic dynamics of Eq.~(\ref{eq:01}) and the underlying network topology remains a central area of research in the nonlinear dynamics and network science communities~\cite{Moreno2004EPL,Ichinomiya2004PRE,Restrepo2005PRE,Arenas2006PRL,GomezGardenes2007PRL,Gardenes2011PRL,Restrepo2014EPL,Skardal2015PRE,Skardal2014PRL}. Recently, the dimensionality reduction method discovered by Ott and Antonsen~\cite{Ott2008Chaos,Ott2009Chaos} has facilitated the analytical treatment of variants of Eq.~(\ref{eq:01}). Examples of the application of the so-called Ott-Antonsen ansatz include external forcing~\cite{Childs2008Chaos}, time-delayed coupling~\cite{Lee2009PRL}, higher-order coupling~\cite{Skardal2011PRE}, adaptive coupling~\cite{Skardal2014PhysicaD}, positive and negative coupling~\cite{Hong2011PRL}, other sinusoidally-coupled phase oscillator systems~\cite{Pazo2014PRX,Laing2014PRE}, and different natural frequency distributions~\cite{Skardal2018PRE}.

One class of oscillator systems for which the Ott-Antonsen ansatz has proven very useful are mean-field approximations of networks with community structure~\cite{Abrams2008PRL,Barreto2008PRE,Laing2009Chaos,Skardal2012PRE,Alonso2011Chaos,Martens2016Chaos,BickPreprint,Pietras2016PRE}, where oscillators are partitioned into groups where coupling between oscillators in the same group differs from that between oscillators in different groups. In different contexts these works reveal the emergence of chimera states, hierarchical path to synchronization, and complex nonlinear behavior. Another important system for which the Ott-Antonsen ansatz allows for analytical treatment that remained elusive is the case of bimodally-distributed frequencies~\cite{Martens2009PRE,Crawford1994JSP}. In this case oscillators tend to organize into ``dynamical'' communities, one for each part of the frequency distribution. In addition to incoherence and partial synchronization, this system displays a ``standing-wave'' state consisting of a limit cycle where oscillators coalesce into two ``giant oscillators'' that does not synchronize, resulting in sustained oscillations in the macroscopic dynamics. In the majority of works studying the dynamics of communities of coupled oscillators particular symmetries are utilized in order to obtain simplified systems that are analytically tractable. The lack of substantial investigations into breaking these symmetries not only represents a gap in our understanding of real systems where such asymmetries may abound, but also impedes our ability to compare the effects of symmetry and symmetry-breaking in coupled oscillator systems.

In this work we study the collective dynamics of a coupled oscillator system with community structure where each community has its own natural frequency distribution. In particular, we consider a system consisting of two communities which gives rise to a possibly bimodal frequency distribution. We begin by analyzing the symmetric case where the properties of the two oscillator communities are equivalent, i.e., their respective natural frequency distributions are the same, save for a difference in their respective means. We then break this symmetry by allowing the respective natural frequency distributions to have different widths, effectively making one community more or less disordered than the other. In the symmetric case we use analytical tools to derive characterize the bifurcation diagram of the system as a function of the intra- and inter-community coupling strength. Interestingly, as long as the mean frequencies of the respective communities are not identical, bistability emerges in the bifurcation diagram between incoherent and partially synchronized states as well as traveling-wave and partially synchronized states. As the difference in the means of the natural frequency distribution of the respective communities is increased, these bistability regions are made more prominent. Moreover, our bifurcation analysis allows us to characterize the transitions between these states and identify a critical value for the difference in natural frequency means that informs the sub- or super-criticality in the transition to partial synchronization. When the symmetry between the two communities find that the macroscopic dynamics become more complicated and more intricate. For instance, while the loss of stability of the incoherent state in the symmetric case came in two types via Hopf and pitchfork bifurcations, respectively, in the asymmetric case this always occurs via a Hopf bifurcation. Moreover, this Hipf bifurcation does not give rise to a traveling-wave state, but rather a partially synchronized state, which can then in turn give rise to a traveling wave state. We also uncover an additional bistability region where two different partially synchronized states, one representing a strongly synchronized state and the other a weakly synchronized state, are both simultaneously stable. Overall, by breaking the symmetry of the system via the communities' natural frequency distributions we uncover a more complicated portrait of the collective dynamics. 

The remainder of this paper is organized as follows. In Sect.~\ref{sec:02} we present the governing equations for the system and as well as the reduced equations for the macroscopic dynamics. In Sect.~\ref{sec:03} we present a bifurcation analysis for the macroscopic system dynamics of the symmetric case. In Sect.~\ref{sec:04} we investigate the properties of the bifurcation diagram for the symmetric case, in particular how its structure and the overall system dynamics change depending on the properties of the natural frequency distributions. In Sect.~\ref{sec:05} we turn our attention to the asymmetric case, where we explore the emergence of a more complicated bifurcation diagram and compare the symmetric and asymmetric cases. In Sect.~\ref{sec:06} we conclude with a discussion of our results.

\section{Governing Equations and Reduced System}\label{sec:02}

\begin{figure}[t]
\centering
\epsfig{file =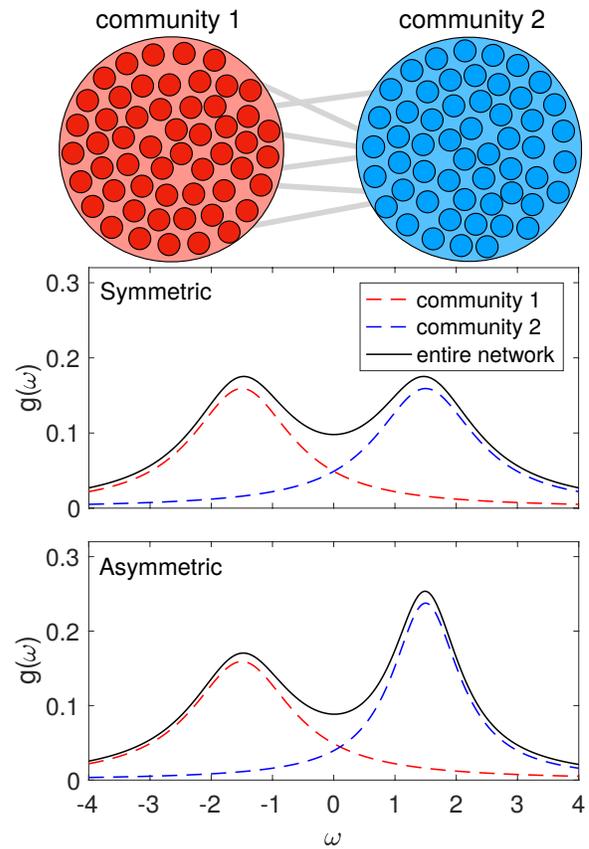, clip =,width=0.95\linewidth }
\caption{(Colour online) {\it Two oscillator communities.} Illustration of the system with two communities of coupled oscillators, along with their respective natural frequency distributions. Illustrated are the symmetric and asymmetric cases in the top and bottom panels, respectively.} \label{fig1}
\end{figure}

In this work we will focus on the case of two interacting groups of coupled phase oscillators. Denoting these groups as $\sigma=1,2$, we assume that both communities are of equal size with $N$ oscillators and consider the following governing equations:
\begin{align}
\dot{\theta}_n^{\sigma}=\omega_n^\sigma+\frac{1}{2}\sum_{\sigma'=1}^2\frac{K^{\sigma\sigma'}}{N}\sum_{m=1}^{N}\sin(\theta_{m}^{\sigma'}-\theta_n^\sigma),\label{eq:02}
\end{align}
where $\theta_n^\sigma$ denotes the phase of oscillator $n$ in community $\sigma$, $\omega_n^\sigma$ is its natural frequency, and $K^{\sigma\sigma'}$ is the coupling strength between oscillators in communities $\sigma$ and $\sigma'$, respectively. We note that these equations of motion are similar to those studied in other works investigating the dynamics of communities of coupled oscillators~\cite{Ott2008Chaos,Abrams2008PRL,Barreto2008PRE,Laing2009Chaos,Skardal2012PRE,Alonso2011Chaos,Martens2016Chaos,BickPreprint,Pietras2016PRE}. Importantly, we will assume that the local dynamics of each oscillator depends on the group to which they belong: the natural frequency of oscillators in community $\sigma$ are drawn from a distribution $g_\sigma(\omega)$ specific to group $\sigma$. In particular, we let $g_\sigma(\omega)$ be Lorenztian with width $\Delta_\sigma$ and mean $\Omega_\sigma$: $g_\sigma(\omega)=\Delta_\sigma/\{\pi[\Delta_\sigma^2+(\omega-\Omega_\sigma)^2]\}$. Note that the overall distribution of natural frequencies, given by $g(\omega)=[g_1(\omega)+g_2(\omega)]/2$ consists of two Lorentzians with possibly different widths and means. The structure of the system is illustrated in Fig.~\ref{fig1}, where each structural group corresponds to a different frequency distribution. As we will see, the dynamics of the system depend on two important properties of the distribution: the separation between the two community sub-distributions and the relative widths of the two community sub-distributions. We note that, by entering a rotating reference frame, we may shift $\Omega_1$ and $\Omega_2$ by the same amount, so without loss of generality we set $\Omega_2=\Omega=-\Omega_1$. Next we consider two different cases of widths. In the symmetric case (top panel), we have that $\Delta_1=\Delta_2=\Delta$, resulting in a symmetric frequency distribution. In the asymmetric case (bottom panel), we have that $\Delta_1\ne\Delta_2$, resulting in an asymmetric frequency distribution. Compared to the symmetric case, the asymmetric Lorentzian cases represents the generalization where oscillator's internal dynamics (i.e., natural frequencies) are more or less heterogeneous in one community compared to the other. Importantly, note that by varying $\Omega$ we may modify the modality of the overall distribution: if $\Omega$ is large enough compared to the widths $\Delta_1$ and $\Delta_2$, the overall frequency distribution is bimodal, whereas if $\Omega$ is too small the overall distribution is unimodal. Finally, to measure the degree of synchronization within each group, we use the following community-wise order parameters
\begin{align}
z_\sigma=r_\sigma e^{i\psi_{\sigma}}=\frac{1}{N}\sum_{m=1}^{N}e^{i\theta_m^\sigma},\label{eq:03}
\end{align}
where the amplitude $r_\sigma$ measures the local degree of synchronization among oscillators in community $\sigma$ and $\psi_\sigma$ gives the mean phase of oscillators in community $\sigma$.

To obtain an analytically tractable system from Eq.~(\ref{eq:02}) we consider the continuum limit of $N\to\infty$ and use Ott and Antonsen's dimensionality reduction. For the case of Lorentzian distributions, this technique is outlined, for instance, in Ref.~\cite{Skardal2012PRE} and results in the following closed-form system consisting of two complex-valued ODEs, one for each of the local order parameters:
\begin{align}
\dot{z}_\sigma = (i\Omega_\sigma-\Delta_\sigma)z_\sigma + \frac{1}{4}\sum_{\sigma'=1}^2K^{\sigma\sigma'}(z_{\sigma'}-z^*_{\sigma'}z_{\sigma}^2),\label{eq:04}
\end{align}
where $^*$ denotes the complex conjugate. Next, we will denote the coupling within each group and between groups as $k$ and $K$, respectively, so that $K^{11}=K^{22}=k$ and $K^{12}=K^{21}=K$. Converting the complex-valued dynamics of Eq.~(\ref{eq:04}) to polar coordinates and introducing the phase difference variable $\psi=\psi_2-\psi_1$ yields the following system of three real-values equations:
\begin{align}
\dot{r}_1 &=-\Delta_1 r_1+\frac{1-r_1^2}{4}\left(kr_1+Kr_2\cos\psi\right),\label{eq:05}\\
\dot{r}_2 &=-\Delta_2 r_2+\frac{1-r_2^2}{4}\left(Kr_1\cos\psi+kr_2\right),\label{eq:06}\\
\dot{\psi} &=2\Omega-\frac{K\left[r_1^2(1+r_2^2)+r_2^2(1+r_1^2)\right]}{4r_1r_2}\sin\psi,\label{eq:07}
\end{align}
Equations~(\ref{eq:05})--(\ref{eq:07}) thus describe the macroscopic system dynamics of the two groups via the amplitudes of the respective order parameters and the relative phase difference. 

\section{Bifurcation Analysis in the Symmetric Case}\label{sec:03}

We begin our analysis by considering the symmetric case and proceed with a bifurcation analysis of the macroscopic dynamics given by Eqs.~(\ref{eq:05})--(\ref{eq:07}) for the choice $\Delta_1=\Delta_2=\Delta$. This system combines features from a number of other oscillator systems, most notably community structure and bimodality in the natural frequency distribution~\cite{Barreto2008PRE,Skardal2012PRE,Martens2009PRE,Pietras2016PRE}, and therefore the bifurcation analysis below will draw on features from these similar cases. Next, we note that Eqs.~(\ref{eq:05})-(\ref{eq:07}) have the symmetry $(r_1,r_2,\psi)\mapsto(r_2,r_1,\psi)$ as well as the invariant manifold defined by $r_1=r_2$. In Ref.~\cite{Martens2009PRE} this manifold was found to be stable when coupling strengths are uniform throughout the system (i.e., $k=K$), and simulations of Eqs.~(\ref{eq:05})--(\ref{eq:07}) suggest that it remains stable in the more general case $k\ne K$ that we study here. Therefore, we will search for the dynamics on this manifold, letting $r_1=r_2=r$. Next we rescale the system to eliminate the parameter $\Delta$ by defining $\tilde{t}=\Delta t/2$, $\tilde{\Omega}=4\Omega/\Delta$, $\tilde{k}=k/\Delta$, and $\tilde{K}=K/\Delta$. Moreover, to simplify the analysis we borrow a technique from Ref.~\cite{Martens2009PRE} and introducing the quantity $q=r^2$. Letting the overdot now denote the derivative with respect to time and dropping the $\sim$ notation for simplicity yields the new system
\begin{align}
\dot{q}&=q[k-4-kq+(1-q)K\cos\psi],\label{eq:08}\\
\dot{\psi}&=\Omega - K(1+q)\sin\psi,\label{eq:09}
\end{align}
where the parameters of interest are (all rescaled) the intra-community coupling strength $k$, the inter-community coupling strength $K$, and the characteristic natural frequency parameter $\Omega$. We now search for solutions representing incoherent, standing-wave, and partially synchronized states characterized, respectively, by the $q=0$ fixed point, a limit-cycle, and a $q>0$ fixed point.

First we consider the stability of the incoherent state, for which case it is most convenient to re-examine the system in Cartesian coordinates, i.e., the real and imaginary parts, of Eq.~(\ref{eq:04}). Examining the eigenvalues of the Jacobian linearized about the state $z_1=z_2=0$, we find that the the incoherent state becomes unstable when the real part of $k-4+\sqrt{K^2-\Omega^2}$ vanishes. This leads to either a a transcritical bifurcation in $q$ or a Hopf bifurcation, depending on the parameters $K$ and $\Omega$. We note that the transcritical bifurcation in $q$ corresponds to a pitchfork bifurcation in $r$, so we will refer to it from here on forward as a pitchfork bifurcation. In terms of the rescaled parameters we have that bifurcation of the incoherent state occur at
\begin{align}
K^2-(k-4)^2&=\Omega^2 \hskip1ex\text{if}\hskip1ex\Omega\le K\hskip2ex\text{(Pitchfork), or}\label{eq:10}\\
k &= 4\hskip1ex\text{if}\hskip1ex\Omega>K\hskip2ex\text{(Hopf)}.\label{eq:11}
\end{align}

Next we turn our attention to the formation and behavior of partially synchronized states characterized by non-zero fixed point solutions of Eqs.~(\ref{eq:08}) and (\ref{eq:09}). Eliminating the $q=0$ solution and the phase parameter $\psi$ using $\sin^2\psi+\cos^2\psi=1$, we rearrange to obtain
\begin{align}
K^2=\frac{\Omega^2}{(1+q)^2}+\frac{(4-k+kq)^2}{(1-q)^2}.\label{eq:12}
\end{align}
First, it is notable that taking the limit $q\to0^+$ allows us to recover precisely the pitchfork bifurcation curve given in Eq.~(\ref{eq:10}). Next, it can be easily verified that for various choices of $k$ and $\Omega$, $K$ is not a monotonic function of $q$, indicating that $q$ is multi-valued, in turn suggesting that the partially synchronized state (one stable, another unstable) is born from a saddle-node bifurcation. When this is the case the saddle-node can be found by enforcing the condition $\partial K/\partial q=0$, yielding
\begin{align}
0=4[4+k(q-1)](1+q)^3+(q-1)^2\Omega^2.\label{eq:13}
\end{align}
Using Eq.~(\ref{eq:12}) to eliminate $q$ in Eq.~(\ref{eq:13}) yields the saddle-node bifurcation curve as a root of the following implicit equation:
\begin{widetext}
\begin{align}
0 &= 64K^8-K^2[4(k-2)^2+\Omega^2]\left[16(k-2)^4+18K^4+16\Omega^2(k^2+7k+2)+\Omega^4\right]\nonumber\\
&+k^2\Omega^2[16(k-2)^4+8\Omega^2(k^2+12k+4)+\Omega^4]+4K^4[48(k-2)^4+4\Omega^2(7k^2+12k-84)+3\Omega^4]\hskip1ex\text{(Saddle-node)}\label{eq:14}
\end{align}
\end{widetext}
While an explicit expression for the saddle-node curve in $(K,k)$ space is difficult to obtain from Eq.~(\ref{eq:14}), we note that roots can easily be obtained numerically using a Newton iteration or other root-finding techniques. Moreover, the saddle-node bifurcation is itself born at a codimension-two point at the intersection of the pitchfork curve [Eq.~(\ref{eq:10})] and saddle-node curve [Eq.~(\ref{eq:14})], indicating the first point where the solutions $q$ first become double-valued, folding onto itself. Inserting Eq.~(\ref{eq:12}) into Eq.~(\ref{eq:14}) yields that this codimension-two point occurs at 
\begin{align}
(K,k)=\left(\Omega\sqrt{1+\frac{\Omega^2}{16}},4-\frac{\Omega^2}{4}\right)\label{eq:15}
\end{align}

\begin{figure*}[t]
\centering
\epsfig{file =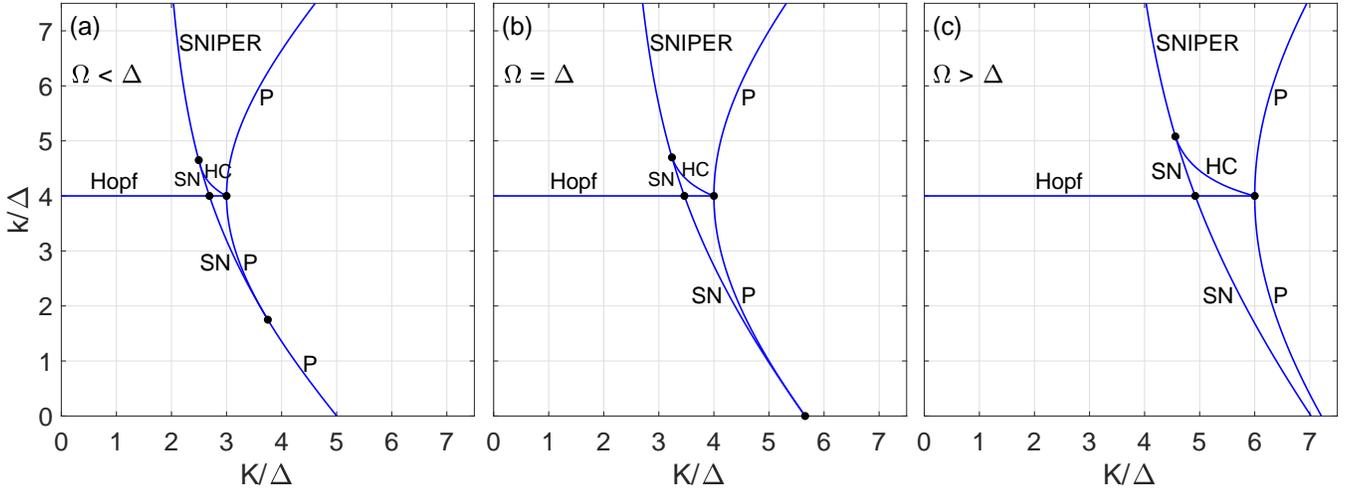, clip =,width=1.0\linewidth }
\caption{(Colour online) {\it Symmetric Case: Bifurcation Diagrams.} Bifurcation diagrams for the macroscopic dynamics given in Eqs.~(\ref{eq:05})--(\ref{eq:07}) in terms of the original system parameters $K/\Delta$ and $k/\Delta$ for three distinct cases: (a) $\Omega<\Delta$, (b) $\Omega=\Delta$, and (c) $\Omega>\Delta$. (Specific values used are $\Omega=3\Delta/4$, $\Delta$, and $3\Delta/2$, respectively.) Bifurcation curves are labeled pitchfork (P), Hopf (Hopf), saddle-node (SN), homoclinic (HC), and SNIPER (SNIPER).} \label{fig2}
\end{figure*}

Lastly, the local bifurcation analysis above does not capture a vital global bifurcation: a homoclinic bifurcation that occurs when the limit-cycle solution born from the Hopf bifurcation collides and annihilates with the unstable fixed point $q>0$. As we will see below, this bifurcation marks the boundary of the bistability region with the standing-wave and partially synchronized states. Due to the global nature of this homoclinic bifurcation, we are required to proceed numerically, and present it in the following section. (To find this bifurcation curve we track track the traveling wave solution as $K$ is increased until it annihilates, which occurs at the homoclinic bifurcation.) However, we will see that this homoclinic bifurcation stretches from the intersection between the pitchfork and Hopf curves to a location along the saddle-node curve. Beyond this point, the saddle-node curve actually corresponds to a saddle-node infinite-period (SNIPER) bifurcation, as the fixed point $q>0$ is born on the limit cycle, resulting in a closed loop that takes an infinitely long time to fully traverse.

\section{Bifurcation Diagrams and Multistability in the Symmetric Case}\label{sec:04}

The results presented above give the full description of the bifurcations that occur between macroscopic states as a function of the coupling strengths $k$ and $K$ for the symmetric Lorentzian case, but the bifurcation curves themselves depend also on the value of the frequency parameter $\Omega$. In Fig.~\ref{fig2} we illustrate structure of the bifurcation diagram for, as we will describe below, three distinct cases. We note that Fig.~\ref{fig2} reports parameters in terms of their original non-scaled values. Bifurcation curves are labeled pitchfork (P), Hopf (Hopf), saddle-node (SN), homoclinic (HC), and SNIPER (SNIPER), with intersections illustrated by black circles. Here we will analyze the structure of the bifurcation diagram and overall properties of the dynamics as the frequency parameter $\Omega$ is varied.

Common to all three cases are several key features. First, the Hopf bifurcation [Eq.~(\ref{eq:11})] collides with the pitchfork bifurcation [Eq.~(\ref{eq:10})] at the point $(K,k)=(\Omega,4)$. Moreover, when it exists, the saddle-node bifurcation [Eq.~(\ref{eq:14})] occurs at a smaller value of $k$ than the pitchfork bifurcation and crosses the Hopf curve at $k=4$, having been born at the codimension-two point at the intersection of the saddle-node and pitchfork curves given in Eq.~(\ref{eq:15}). These three bifurcation curves represent the boundary of a region of bistability between the incoherent (i.e., $q=0$ fixed point) and partially synchronized (i.e., $q>0$ fixed point) states. Importantly, we also report that this region of bistability exists for all values of $\Omega\ne0$. This can be seen by inspecting the codimension-two point at the intersection of the pitchfork and saddle-node curves. Specifically, this intersection occurs at $k=4-\Omega^2/4<4$, whereas the Hopf curve is given by precisely $k=4$. Since the saddle-node curve must lie to the left of the pitchfork curve, this yields a bistability region with positive area for any $\Omega\ne0$.

Also common to all three cases illustrated in Fig.~\ref{fig2} is the homoclinic curve that stretches from the intersection of the Hopf and pitchfork curves [at $(K,k)=(\Omega,4)$] to a point along that saddle-node curve beyond the Hopf curve at some value $k>4$. (Recall that this curve is calculated numerically due to the global nature of the bifurcation.) The homoclinic curve completes the boundary of another bistability region enclosed by the Hopf, saddle-node and homoclinic curves. Note that this bistability region lies beyond the Hopf curve, indicating that it represents bistability between the standing-wave (i.e., limit-cycle) and partially synchronized (i.e., $q>0$ fixed point) states. We may also conclude that this bistability region must exist for all $\Omega\ne0$. This follows from the fact that it shares a boundary the previous bistability region, which has positive area, and our numerical simulations suggest that the intersection between the saddle-node and homoclinic curves always occurs at $k>4$, resulting in another bistability region with positive area.

These results reveal that, as long as the difference between frequency distributions is non-zero, i.e., $\Omega\ne0$, the system always presents bistability for some combination of coupling strengths. We note here that this is in contrast to the case of bimodally-distributed frequencies without community structure, where bistability may only occur for a relatively small set of frequency parameters. In particular, to observe bistability without community structure the difference parameter $\Omega$ must be sufficiently large, but not too large, compared to the width parameter $\Delta$ (which we scaled out of the governing equations). This range corresponds to values of $\Omega$ ensuring that the overall distribution is actually bimodal, but has peaks that are sufficiently close. Here we see that, regardless of $\Delta$, any $\Omega\ne0$ allows for bistability. This indicates that the freedom to tune intra- and inter-community coupling strengths differently promotes bistability in the system.

In addition to the presence of bistability for all $\Omega\ne0$, the overall structure of the bifurcation diagram depends critically on $\Omega$. The three different cases of bifurcation diagrams in Fig.~\ref{fig2} correspond to increasing the frequency parameter: (a) $\Omega=3$, (b) $4$, and (c) $6$, respectively. More specifically, an important distinction can be observed between these cases by studying the codimension-two point at the intersection of the pitchfork and saddle-node curves given in Eq.~(\ref{eq:15}). Restricting our attention to non-negative coupling strengths, i.e., the quadrant $K\ge0$, $k\ge0$, we find that this codimension may or may not fall in this quadrant. Specifically, this point falls precisely on the boundary, $k=0$, when $\Omega=4$. Moreover, for $\Omega<4$ this codimension-two point lies within this quadrant, but for $\Omega>4$ it escapes this quadrant. Therefore, a critical difference in the bifurcation diagrams occurs at $\Omega=4$. (In terms of the original non-dimensionalized parameters this is given by $\Omega=\Delta$.) In addition to the layout of the bifurcation diagrams depicted in Fig.~\ref{fig2}, this has an important effect on the dynamics themselves, most notably the transition from incoherence to partial synchronization. For $\Omega<4$ this transition occurs in two flavors. For small enough $k$, specifically $0\le k\le4-\Omega^2/4$, a transition to partial synchronization is made with no hysteresis, owing to a supercritical pitchfork bifurcation. For larger $k$, specifically $4-\Omega^2/4<k\le4$, this transition fold onto itself into a hysteresis loop, owing to a subcritical pitchfork bifurcation. However, for $\Omega>4$ this transition occurs in only one flavor: a hysteresis loop exists for all $0\le k\le4$ owing to a subcritical pitchfork bifurcation.

\begin{figure}[b]
\centering
\epsfig{file =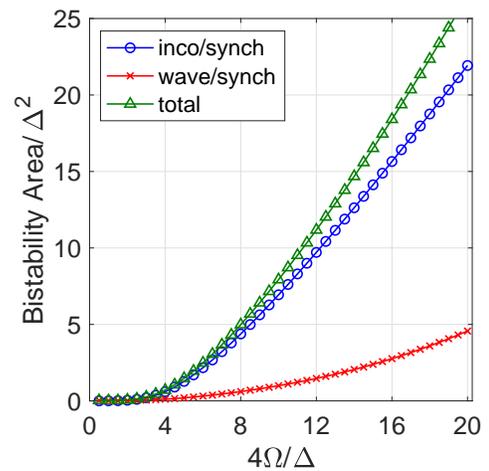, clip =,width=0.8\linewidth }
\caption{(Colour online) {\it Bistability Region Areas.} Areas in $(K,k)$ parameter space corresponding to bistability between the incoherent/partially synchronized states (blue circles) and the standing-wave/partially synchronized states (red crosses) ad a function of $4\Omega/\Delta$. Total combined bistability area is plotted in green triangles.} \label{fig3}
\end{figure}

Next, we investigate the degree to which the difference between the natural frequency distributions promotes bistability by calculating the area of the respective bistability regions as $\Omega$ is varied. In Fig.~\ref{fig3} we plot the numerically calculated area of each bistability region, denoting the incoherent/partially synchronized region with blue circles and the standing-wave/partially synchronized region with red crosses, as well as their combined area with green triangles. Results are reported in terms of the original system parameters, noting that the area of a region in $\left(K,k\right)$ space is $\Delta^2$ times the corresponding area in $\left(K/\Delta,k/\Delta\right)$ space. While the area of both bistability regions increase with $\Omega$, we observe that the area of the incoherent/partially synchronized bistability region contributes significantly more than the standing-wave/partially synchronized bistability region.

Lastly, we consider the behavior of the system in the case of strong community structure, i.e., the limit $k\to\infty$. In this regime we may see in Fig.~\ref{fig2} that the system may display either standing-wave or partially synchronized behaviors, but not incoherence. Note also that for sufficiently large $k$ we surpass the intersection of the saddle-node and homoclinic curves, so the transition from a standing-wave to partial synchronization occurs via a SNIPER bifurcation. From Eqs.~(\ref{eq:05}) and (\ref{eq:06}) we see that for $k\gg1$ and $\Delta,K\sim1$ we have that $r_1,r_2\approx1$. Noting that $r_1,r_2=1$ corresponds to $q=1$, we find that in the limit $k\to\infty$ the phase-difference dynamics are given (in rescaled parameters) by
\begin{align}
\dot{\psi}=\Omega-2K\sin\psi.\label{eq:16}
\end{align}
This yields a transition from standing-waves to partial synchronization at
\begin{align}
K_\infty=\frac{\Omega}{2},\label{eq:17}
\end{align}
which matches with our numerical simulations (not shown).

\section{Collective Dynamics in the Asymmetric Case}\label{sec:05}

We now turn our attention to the asymmetric case where the widths of the natural frequencies distributions for the respective communities are unequal, i.e., $\Delta_1 \ne \Delta_2$. Since one community's width must be larger than that of the other we assume without loss of generality that $\Delta_1>\Delta_2$. (Note that if $\Delta_2>\Delta_1$ we may simply rename the two community assignments.) We find it convenient then to rename $\Delta_1=\Delta_{\text{max}}$ and $\Delta_2=\Delta_{\text{min}}$. This generalization of the system studied in the previous two sections breaks a critical symmetry in the system, which we will see leads to more complicated dynamics that emerge. 

\begin{figure}[t]
\centering
\epsfig{file =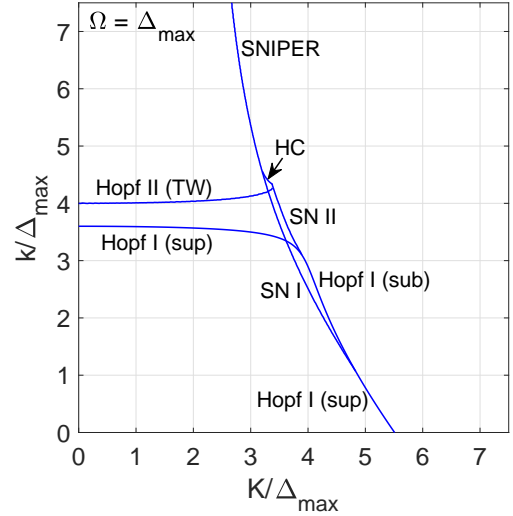, clip =,width=0.9\linewidth }
\caption{(Colour online) {\it Asymmetric Case: Bifurcation Diagram.} Bifurcation diagrams for the macroscopic dynamics given in Eqs.~(\ref{eq:05})--(\ref{eq:07}) in terms of the original system parameters $K/\Delta_{\text{max}}$ and $k/\Delta_{\text{max}}$ for the case of $\Omega = \Delta_{\text{max}}$ with $\widehat{\Delta}=0.9$. Bifurcation curves are labelled as described in the main text.} \label{fig4}
\end{figure}

We begin by considering a similar rescaling of the system as before, now scaling the larger width, i.e., the width of the first community, $\Delta_{\text{max}}$ out of the equations of motion by defining $\tilde{t}=\Delta_{\text{max}} t/2$, $\tilde{\Omega}=4\Omega/\Delta_{\text{max}}$, $\tilde{k}=k/\Delta_{\text{max}}$, $\tilde{K}=K/\Delta_{\text{max}}$, and $\widehat{\Delta}=\Delta_{\text{main}}/\Delta_{\text{max}}$, which yields (after dropping the $\sim$ notation again)
\begin{align}
\dot{r}_1 &=-2 r_1+\frac{1-r_1^2}{2}\left(kr_1+Kr_2\cos\psi\right),\label{eq:18}\\
\dot{r}_2 &=-2\widehat{\Delta}r_2+\frac{1-r_2^2}{2}\left(Kr_1\cos\psi+kr_2\right),\label{eq:19}\\
\dot{\psi} &=\Omega-\frac{K\left[r_1^2(1+r_2^2)+r_2^2(1+r_1^2)\right]}{2r_1r_2}\sin\psi.\label{eq:20}
\end{align}
In particular, the new parameter $\widehat{\Delta}=\Delta_{\text{min}}/\Delta_{\text{max}}$ represents the non-dimensional ratio of the smaller width to the larger width, so that $0<\widehat{\Delta}<1$. Ths breaks the symmetry between Eqs.~(\ref{eq:14}) and (\ref{eq:15}), suggesting that in a non-incoherent state $r_1$ and $r_2$ are unlikely to be equal to one another, so we may not reduce the three dimensional system given by Eq.~(\ref{eq:18})--(\ref{eq:20}) to a two dimensional system as we did in the symmetric case [i.e., Eqs.~(\ref{eq:08}) and (\ref{eq:09})].

\begin{figure*}[t]
\centering
\epsfig{file =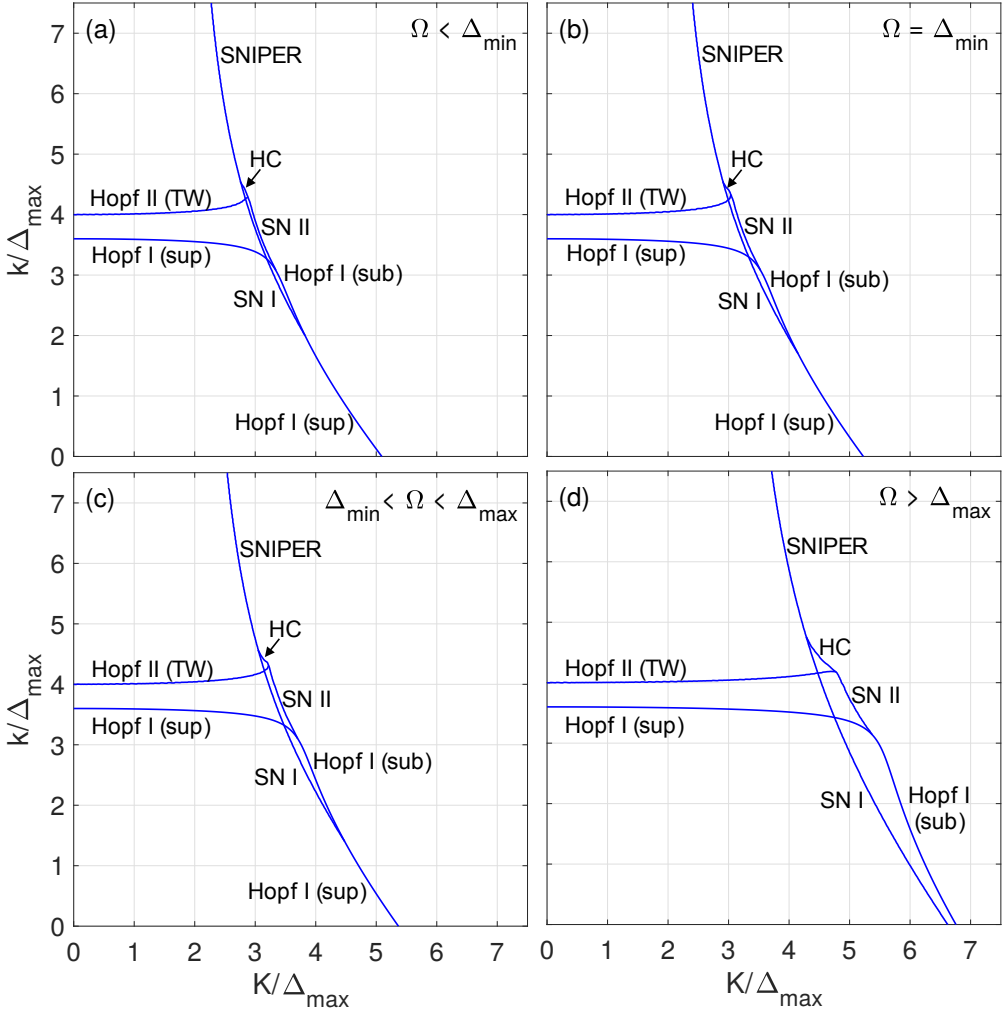, clip =,width=0.8\linewidth }
\caption{(Colour online) {\it Asymmetric Case: Bifurcation Diagrams (Others).} Bifurcation diagrams for the macroscopic dynamics given in Eqs.~(\ref{eq:05})--(\ref{eq:07}) in terms of the original system parameters $K/\Delta_{\text{max}}$ and $k/\Delta_{\text{max}}$ for the cases: (a) $\Omega < \Delta_{\text{min}}$, (b) $\Omega = \Delta_{\text{min}}$, (c) $\Delta_{\text{min}} < \Omega < \Delta_{\text{max}}$, and (d) $\Omega > \Delta_{\text{max}}$ with $\widehat{\Delta}=0.9$. (Specific values used are $\Omega=0.85\Delta_{\text{max}}$, $\Omega=0.9\Delta_{\text{max}}$, $\Omega=0.95\Delta_{\text{max}}$, and $\Omega=1.4\Delta_{\text{max}}$, respectively.) Bifurcation curves are labelled as described in the main text.} \label{fig5}
\end{figure*}

Next we consider the stability of the incoherent state described by $r_1,r_2=0$. As in the symmetric case, this is most conveniently done by examining the system in Cartesian coordinates, i.e., Eq.~(\ref{eq:04}). Examining the Jacobian matrix evaluated at the incoherent state, we find that the eigenvalues are given by (in terms of the rescaled parameters)
\begin{align}
\frac{\lambda}{\Delta_{\text{max}}}&=\frac{k-2(1+\widehat{\Delta})\pm\sqrt{K^2+[2(1-\widehat{\Delta})\pm i\Omega]^2}}{4}.\label{eq:21}
\end{align}
Specifically, the incoherent state loses stability when the eigenvalue(s) with largest real part pass through the imaginary axis from negative real part to positive real part. However, note that since $0<\widehat{\Delta}<1$ the right hand side of Eq.~(\ref{eq:21}) has a non-zero imaginary part. Therefore, the loss of stability of the incoherent state must occur in the form of a Hopf bifurcation, regardless of the parameters. The bifurcation curve is given implicitly in the $(K,k)$ parameter space when the eigenvalue corresponding to the plus sign outside of the square root has exactly zero part and can easily be calculated numerically. However, it is possible to find the bifurcation values in the two limits $K\to0$ and $k\to 0$. Beginning with $K\to0$, some algebraic manipulation (note that $1-\widehat{\Delta}>0$) simplifies Eq.~(\ref{eq:21}) to give a critical coupling strength of $k_{K=0}=4\widehat{\Delta}$ (Hopf I). On the other hand, when $k\to0$ Eq.~(\ref{eq:21}) can be simplified to give a critical coupling strength of $K_{k=0}=\sqrt{16\widehat{\Delta}+\Omega^2}$ (Hopf I).

\begin{figure*}[t]
\centering
\epsfig{file =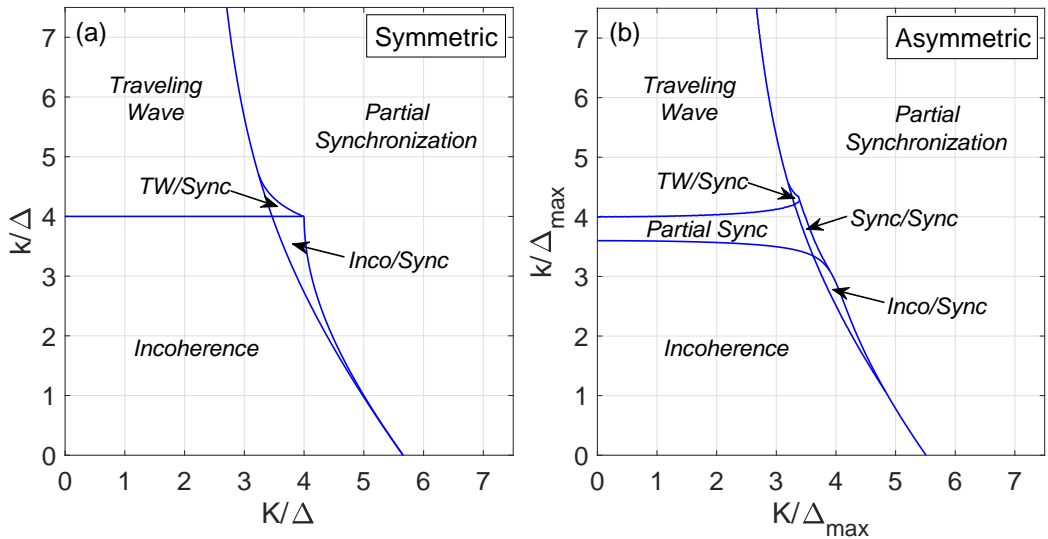, clip =,width=0.8\linewidth }
\caption{(Colour online) {\it Symmetry and symmetry breaking.} Representative stability diagrams for (a) the symmetric and (b) asymmetric cases, using $\Omega=\Delta$ and $\Omega=\Delta_{\text{max}}$, respectively. (In the latter case we choose $\widehat{\Delta}=0.9$.)} \label{fig6}
\end{figure*}

This Hopf bifurcation, which corresponds to the loss of stability of the incoherent state, deserves a few more remarks that differentiates the asymmetric case from the symmetric case. First, it is the first of two Hopf bifurcations that occur in the system. In particular, for $k$ sufficiently large compared to $K$, in the symmetric case the incoherent state gave way directly to the traveling wave state. In the asymmetric case, however, the incoherent state gives way to a partially synchronized state through the first Hopf bifurcation where $r_1$ and $r_2$ approach a constant value and the two communities phase lock, as indicated by $\psi$ reaching a constant value while $\psi_1$ and $\psi_2$ continue to process at a constant rate. Moreover, this Hopf bifurcation (denoted Hopf I, since it is the first of two Hopf bifurcations) occurs earlier (i.e., at smaller $k$) than in the symmetric case because this particular bifurcation corresponds to community $2$ (i.e, that with thinner width) synchronizing earlier than community $1$ but $z_2$ ``dragging'' $z_1$ along. When $k$ is increased further a second Hopf bifurcation (Hopf II) occurs which gives rise to the traveling wave state where, as in the symmetric case, the amplitudes $r_1$ and $r_2$ begin to oscillate. We illustrate this structure in Fig.~\ref{fig4}, plotting the bifurcation diagram for the case $\Omega = 4$ ($\Omega=\Delta_{\text{max}}$ in the non-scaled system parameters) for $\widehat{\Delta}=0.9$. We note here that for the asymmetric case the Hopf I curve is obtained by solving Eq.~(\ref{eq:21}) for $\text{Re}(\lambda)=0$, while other bifurcation curves are obtained by direct numerical simulation, similar to the Homoclinic curve in the symmetric case. We also note that in the region of parameter space where $k$ is sufficiently large compared to $K$ the first Hopf curve is supercritical and the second corresponds to the traveling wave solution, so we label them accordingly. Second, in the other region of parameter space where $K$ is sufficiently large compared to $k$, we again see the emergence of bistability through subcriticality, now through a Hopf bifurcation. In particular, at a large enough value of the Hopf I curve changes from supercritical to subcritical as a saddle node bifurcation (of cycles) is born, labeled SN I because it is also the first of two bifurcations of the same type. The subcriticality in the Hopf bifurcation yields the first bistability region between incoherence and partial synchronization, bounded between the Hopf I and SN I curves.

The asymmetric case then presents a new bistability region, beyond the first bistability region, where two stable partially synchronized states coexist. These two states are precisely the lower synchronization state that emerges from the Hopf I curve in the large $k$ region of parameter space and the greater synchronization state that emerges from the SN I curve. In Fig.~\ref{fig4} this is given by the relatively thin region bounded between the Hopf I and Hopf II curves in the $k$ direction and the SN I and SN II curves in the $K$ direction. This second saddle-node (of cycles) curve represents the annihilation of the lower synchronization state as it collides with an unstable partially synchronized state in a typical saddle-node bifurcation (of cycles). Beyond the new bistability regions lies a third bistability region which, similar to the symmetric case, features the coexistence of the traveling wave state and the (greater) synchronization state. This bistability region is bounded by the Hopf II curve, the SN I curve, as well as a Homoclinic curve, beyond which the SN I bifurcation becomes a SNIPER bifurcation.

In Fig.~\ref{fig4} we have plotted the bifurcation diagram for the specific case of (in terms of the original unscaled parameters) $\Omega = \Delta_{\text{max}}$. We also explore the other possibilities, plotting in Fig.~\ref{fig5} the bifurcation diagrams for the four other cases,  $\Omega < 4\widehat{\Delta}$, $\Omega = 4\widehat{\Delta}$, $4\widehat{\Delta} < \Omega < 4$, and $\Omega > 4$ in panels (a)--(d), respectively. In particular, this allows us to observe how the shape and size of the overall bistability region changes with respect to varying parameters. As in the symmetric case, we can see plainly that increasing $\Omega$ results in a larger region of bistability. However, it is also interesting to note that by decreasing $\widehat{\Delta}$ (i.e., increasing the difference in $\Delta_1$ and $\Delta_2$), the bistability range shrinks. Finally, from these numerical investigations we observe that the codimension-two point at the intersection of the saddle-node bifurcation of cycles and the Hopf bifurcation escapes outside of the positive $(K,k)$ parameter space for some $\Omega>4$. [Numerical simulations (not shown) indicate that the precise value of $\Omega$ at which this occurs depends on $\widehat{\Delta}$.] This is in contrast to the symmetric case when this occurs at precisely $\Omega = 4$.

Lastly, we seek to compare the overall behavior of the asymmetric case to the symmetric case. In Fig.~\ref{fig6} we plot the stability diagrams for the representative cases of the (a) symmetric and (b) asymmetric cases for $\Omega=4$. Instead of labeling bifurcation curves we label the regions of parameter space where the different solutions are stable. In the Symmetric case we observe incoherence, traveling wave, and partial synchronization along with bistable incoherence/partial synchronization and traveling wave/partial synchronization states. In the asymmetric case we observe the same states along with an addition region of parameter space corresponding to partial synchronization and a bistable partial synchronization/partial synchronization state. Comparing the results plotted in Fig.~\ref{fig6} (a) and (b) allows us to see the increase in complexity that arises in the collective dynamics of the system when symmetry is broken.

\section{Discussion}\label{sec:06}

In this paper we have studied the effects of symmetry and symmetry breaking on the collective dynamics of two interacting communities of coupled phase oscillators with different natural frequency distributions. First we have used analytical tools to derive the bifurcation diagram for the symmetric case, where the widths of the natural frequency distributions for the respective natural frequency distributions are equal, indicating an equal degree of disorder in the respective communities' internal dynamics. We then break this symmetry by allowing the two frequency widths to be unequal, so that one community's internal dynamics is more or less disordered than the other. 

In the symmetric case we find dynamical states corresponding to incoherence, standing-waves, and partial synchronization, as well as two regions of bistability between incoherence and partial synchronization and between traveling waves and partial synchronization. Moreover, the freedom to vary intra- and inter-community coupling strengths allows for bistability to occur for all choices of natural frequency distributions provided that the distributions for the respective communities are not identical. This is in contrast to the system without community structure, in which case the frequency distribution must be bimodal, but with sufficiently close peaks to observe bistability. Moreover, by studying the properties of a codimension-two point, we find that a critical value of the natural frequency distribution difference exists. Below this critical value the transition from incoherence to partial synchronization may come in either a supercritical or subcritical pitchfork bifurcation, with the latter case yielding a hysteresis loop. However, above this critical value the transition invariably occurs via a subcritical pitchfork bifurcation.

In the asymmetric case we find that the collective dynamics become more complicated, resulting in a more intricate bifurcation diagram. First, the asymmetry in the system breaks the pitchfork bifurcation, so that the loss of stability of incoherence always comes via a Hopf bifurcation. Second, the incoherent state never gives way to the traveling wave state. Rather, in the appropriate region in parameter space a thin layer exists where a second partially synchronized state exists, which in turn gives way to the traveling waves. Finally, a new bistability region emerges between two different partially synchronized states corresponding to weaker and stronger synchronization, respectively.

Overall, our results highlight the complicating effect that breaking symmetries may have on the collective dynamics of oscillator systems. Here we have focused on breaking the symmetry of the system via the widths of the natural frequency distributions of the respective communities, which corresponds to more or less disorder in the internal dynamics of the oscillators in different communities. However, other methods of symmetry breaking may be found to have different effects. For instance, one may vary the relative sizes of the communities, the intra- or inter-community coupling strengths, or even the coupling functions themselves. Like many other works, we have also focused on the case of two communities here to obtain a more analytically tractable system, however introducing more communities or coupling patterns certainly complicate the dynamics further. In such cases analytical results likely unattainable, however, as we demonstrate in this work, numerical methods and numerical simulations can be successfully used to gain some understanding of the underlying system dynamics, allowing for possible progress in more realistic scenarios. 



\bibliographystyle{plain}

\end{document}